\documentclass{article}
\usepackage[T1]{fontenc} %
\usepackage[utf8]{inputenc} %
\usepackage{ismir,amsmath,cite,url}
\usepackage{graphicx}
\usepackage{color}

\usepackage{times}
\usepackage{soul}
\usepackage{url}
\usepackage[hidelinks,bookmarks=false]{hyperref}
\usepackage[small]{caption}
\usepackage{amsfonts}
\usepackage{amsthm}
\usepackage{booktabs}
\usepackage{algorithm}
\usepackage{algorithmic}
\usepackage[switch]{lineno}
\usepackage[labelformat=simple]{subcaption}
\usepackage{tabularx}
\usepackage{enumitem}
\usepackage{multirow}
\usepackage[table,xcdraw]{xcolor}

\usepackage{musicography}

\usepackage{lineno}

\title{Polyffusion: A Diffusion Model for Polyphonic Score Generation with Internal and External Controls}

\multauthor
{Lejun Min$^{1,2,4}$ \hspace{1cm} Junyan Jiang$^{1,2}$ \hspace{1cm} Gus Xia$^{1,2}$ \hspace{1cm} Jingwei Zhao$^{3}$}
{
$^1$ Music X Lab, Computer Science Department, NYU Shanghai\\
$^2$ MBZUAI \quad $^3$ Institute of Data Science, NUS\\
$^4$ Zhiyuan College, Shanghai Jiao Tong University\\
{\tt\small \href{mailto:aik2mlj@gmail.com}{aik2mlj@gmail.com}, \{\href{mailto:jj2731@nyu.edu}{jj2731}, \href{mailto:gxia@nyu.edu}{gxia}\}@nyu.edu, \href{mailto:jzhao@u.nus.edu}{jzhao@u.nus.edu}}
}

\def\authorname{L. Min, J. Jiang, G. Xia, and J. Zhao}

\begin{document}

\maketitle
\begin{abstract}
We propose Polyffusion, a diffusion model that generates polyphonic music scores by regarding music as image-like piano roll representations. The model is capable of controllable music generation with two paradigms: \textit{internal} control and \textit{external} control. Internal control refers to the process in which users pre-define a part of the music and then let the model infill the rest, similar to the task of masked music generation (or music inpainting). External control conditions the model with external yet related information, such as chord, texture, or other features, via the cross-attention mechanism. We show that by using internal and external controls, Polyffusion unifies a wide range of music creation tasks, including melody generation given accompaniment, accompaniment generation given melody, arbitrary music segment inpainting, and music arrangement given chords or textures. Experimental results show that our model significantly outperforms existing Transformer and sampling-based baselines, and using pre-trained disentangled representations as external conditions yields more effective controls.\footnote{Demo page: \url{https://polyffusion.github.io/}. Code repository: \url{https://github.com/aik2mlj/polyffusion}}
\end{abstract}

\section{Introduction}

Diffusion models~\cite{sohl2015deep,ho2020denoising}, as a new class of generative models, have been successful in generating high-quality samples of image data and beyond. They achieve state-of-the-art sample quality on a number of image generation benchmarks~\cite{dhariwal2021diffusion,ho2022cascaded}, and also show strong results for the generation of various media such as audio~\cite{kong2020diffwave,chen2020wavegrad}, video~\cite{yang2022diffusion,ho2022video,harvey2022flexible}, and text~\cite{li2022diffusion,gong2022diffuseq}.

\begin{figure}[t]
    \centering
    \includegraphics[width=0.45\textwidth]{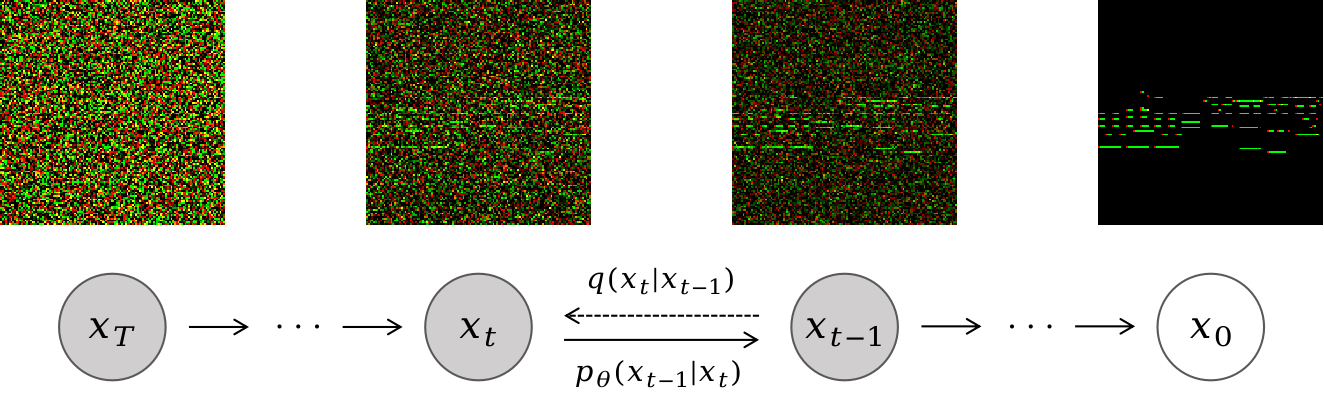}
    \caption{The forward and reverse process of the proposed diffusion model trained on piano roll representations.  The red dot at the front of each note denotes its onset; the green bar following it denotes its sustain. Notice that the image axes are swapped for proper visualization.}
    \label{fig:diffusion}
\end{figure}

Symbolic music generation, a task very different from audio generation, has highly discrete outputs and is often described in terms of constraint optimization problems \cite{pachet2001musical,wang2020learning}. Despite the improvement of deep music generative modeling~\cite{jukebox, musiclm}, symbolic music generation still suffers from the lack of controllability and consistency at different time scales~\cite{briot2020deep}. In our study, we experiment with the idea of using diffusion models to approach controllable symbolic music generation. 

Inspired by the high-quality and controllable image generation that diffusion models have achieved in computer vision, we devise an image-like piano roll format as the input, and used a UNet-based diffusion model to step-wise denoise a randomly sampled piano roll, as illustrated in Figure~\ref{fig:diffusion}. We show in our experiments and demos that our design provides excellent generation results.

Besides unconditional generation, the model also accepts two categories of controls, namely internal control and external control:
\begin{itemize}[leftmargin=*]
  \item \textbf{Internal Control (Inpainting):} By masking out part of the given piano roll, we can specify the remaining area to be generated, thus implicitly conditioning the generation to fit in the masked part. We regard this strategy as a generalized music inpainting method. %
  \item \textbf{External Control (Conditional Generation):} By adopting the cross-attention mechanism of Latent Diffusion~\cite{rombach2022high}, we can explicitly control the music generation on given external conditions such as chords and textures. They are first encoded into latent representations using pre-trained, disentangled variational autoencoders (VAEs), and then fed into the backbone UNet of the diffusion model to condition the denoising process. We show that the generated music complies with the given conditions well. We also add classifier-free guidance to control the variance of the generation.
\end{itemize}

These controls of diffusion models enable us to unify a wide spectrum of creative music tasks that previously require separate modeling and training. In this paper, we showcase the following scenarios:
\begin{itemize}[leftmargin=*]
    \item \textbf{Melody generation given accompaniment} by generation with the accompaniment part being masked out.
    \item \textbf{Accompaniment generation given melody} by generation with the melody part being masked out.
    \item \textbf{Arbitrary music segment inpainting} by generation with any time segments being masked out.
    \item \textbf{General music arrangement given chords or textures} by conditioning on external chord or texture signals. 
\end{itemize}

\section{Related Work}
We review three realms of related work: 1) music inpainting, which is related to our internal control method, 2) conditioned music generation with external signals, which is related to our external control method, and 3) recent progress on diffusion-based modeling in the music domain.

\subsection{Music Inpainting}
Music inpainting is a controlled music generation task that regulates the generation with pre-defined musical contexts. We see various studies on polyphonic music inpainting. For example, DeepBach~\cite{hadjeres2017deepbach} develops a context-aware recurrent neural network (RNN) capable of inpainting missing notes for chorales in the style of Bach. Coconet~\cite{huang2019counterpoint} uses blocked Gibbs sampling to repeatedly rewrite a masked music score. Chang et al.~\cite{chang2021variable} achieve variable-length music score inpainting. Music SketchNet~\cite{chen2020music} and MusIAC~\cite{guo2022musiac} introduce various controls to the inpainting task under VAE-based and Transformer-based framework respectively. Comparatively, diffusion models naturally possess the inpainting ability via masked generation~\cite{lugmayr2022repaint}, and there is no need to train or fine-tune a task-specific model for inpainting.

Though the current inpainting tasks mostly apply masks over a continuous period of time, the inpainted area, in theory, can be any note in the score (any area of a piano roll). In this study, we show that our image-like representation enables both part-wise and time-wise inpainting. The former refers to inpainting melody or accompaniment part given the other part, while the latter refers to infilling notes falling in arbitrary time segments.

\subsection{Music Generation Conditioned on External Signals}
External control signals are also one of the mainstream methods to control the music generation process. Common scenarios include generating music given chords~\cite{simon2008mysong,huang2020pop,hadjeres2017deepbach,donahue2019lakhnes}, lyrics~\cite{ju2021telemelody}, and other relevant features such as note density and voicing numbers~\cite{zhao2021accomontage}.

Our study focuses on polyphonic score generation controlled by external chords and textures. In particular, the “control by texture” task has great practical value in both music arrangement and composition style transfer~\cite{dai2018music}, while very few existing models could realize this function.

\subsection{Diffusion Models for Music Generation}
Recently, we have seen several attempts to introduce diffusion models to symbolic music tasks. Mittal et al.~\cite{mittal2021symbolicdiffusion} generate monophonic music by training a diffusion model on the latent representations learned by MusicVAE~\cite{roberts2018hierarchical}. Cheuk et al.~\cite{cheuk2022diffroll} brings diffusion models to the music transcription task by adapting the piano roll format into the DiffWave~\cite{kong2020diffwave} structure.  It is relevant to our study as the model can also output piano rolls. However, the model focuses on transcription instead of generation by relying on a ground-truth spectrogram as its control. In general, for symbolic music generation, conditioning diffusion models on external controls is still an area to be explored. %

\section{Methods}

\subsection{Data Representation}
\label{sec:datarep}

Our image-like piano roll representation is a 2-channel \textit{binary} tensor $x\in \mathbb{R}^{2\times T \times P}$. The generation task targets 8-bar (32-beat) long music segments, with 1/4 beat as the time step, resulting in $T=128$ time steps per sample. We use a MIDI pitch range $0...127$, resulting in $P=128$ pitch bins. Each entry $x(c, t, p)$ represents whether there is a note onset (for $c=0$) or sustain (for $c=1$) at time step $t$ and MIDI pitch $p$.  %

\subsection{Diffusion Model}

\begin{figure}[]
  \centering
    \includegraphics[width=0.45\textwidth]{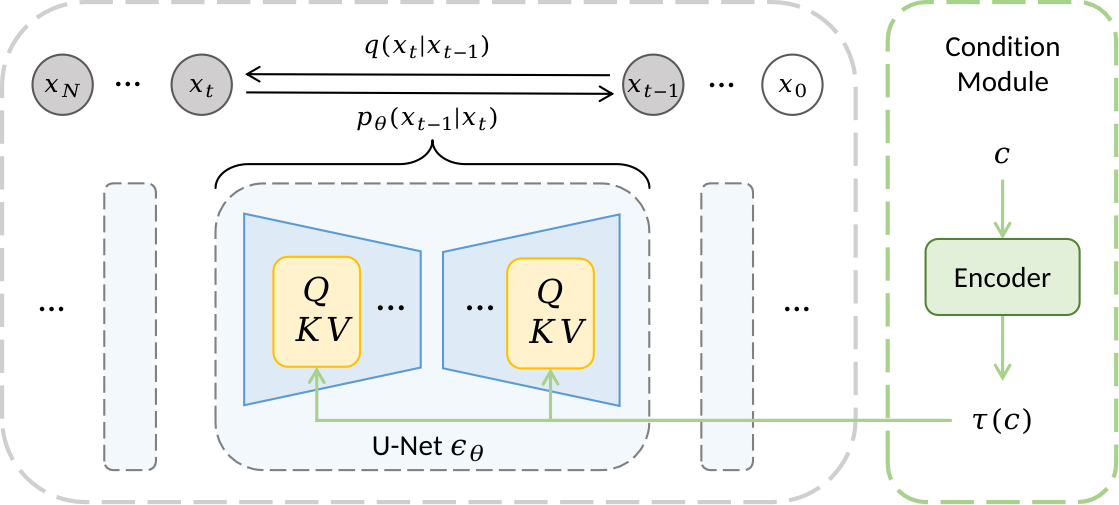}
  \caption{The model structure with an additional condition module for external control. Each UNet unit $\epsilon_\theta$ applies one denoising step during the reverse process. External condition signals are encoded by pre-trained encoders and fed into the cross-attention layers, which are represented by the yellow squares in the UNet unit.}
  \label{fig:model}
\end{figure}

Diffusion models~\cite{sohl2015deep,ho2020denoising} are latent-variable models comprised of a forward (diffusion) process which gradually disrupts the structure of data $x_0$ and a reverse (denoising) process that learns to recover the original data $x_0$ from the noisy input. In our study, $x_0$ denotes the clean piano roll. The forward process iteratively adds Gaussian noise in $N$ diffusion steps:
\begin{equation}
    q(x_t | x_{t-1})=\mathcal{N}(x_t ; \sqrt{1-\beta_t} x_{t-1}, \beta_t I)
\end{equation}
\begin{equation}  
    q(x_{1: N} | x_0)=\prod_{t=1}^N q(x_t | x_{t-1})
\end{equation}
where $\beta_1, \beta_2,\dots, \beta_N$ are a series of variance scheduling parameters.
The reverse process requires the model to parameterize a Markov chain that iteratively reconstructs the piano roll $x_0$ from a corrupted input $x_N\sim \mathcal{N}(0,I)$.
\begin{equation}  
    p_\theta(x_{t-1} | x_t)=\mathcal{N}(x_{t-1} ; \mu_\theta(x_t, t), \sigma_\theta(x_t, t))
\end{equation}
\begin{equation}
    p_\theta(x_{0: N})=p(x_N) \prod_{t=1}^N p_\theta(x_{t-1} | x_t)
\end{equation}

During training, we optimize the model parameters $\epsilon_\theta$ by minimizing the following target:
\begin{equation}
L(\theta)=\mathbb{E}_{x_0, \epsilon, t}\left[\left\|\epsilon-\epsilon_\theta\left(\sqrt{\bar{\alpha}_t} x_0+\sqrt{1-\bar{\alpha}_t} \epsilon, t\right)\right\|^2\right]
\end{equation}
where $t$ is uniformly sampled from $[1,N]$ and $\epsilon\sim \mathcal{N}(0,I)$, $\alpha_t := 1 - \beta_t$, $\bar{\alpha}_t := \prod_{s=1}^t{\alpha_s}$. As shown in Figure~\ref{fig:model}, our unconditional model structure is based on~\cite{ho2020denoising}, an image-oriented diffusion model using a 2-D UNet as its backbone $\epsilon_\theta$.

\subsection{Internal Control (Inpainting)}
Internal control refers to the use of the music notes themselves to regulate and influence the generation process, and we regard music inpainting as a means of internal control. %

Specifically, we denote the given piano roll sample as $s$ and the mask as $m$. At each step $t$ during inference sampling, the fixed area of the image is diffused with the forward process $q(s_t|s) = \mathcal{N}(s_t;\sqrt{\bar{\alpha}_t}s, (1 - \bar{\alpha}_t) I)$ and put together with the denoising sample $s_{t-1}$. Algorithm~\ref{alg:inpaint} ~\cite{lugmayr2022repaint} shows the detailed implementation of this inpainting process.

\begin{algorithm}
\caption{Inpainting Process}
\textbf{Input}: inpainting mask $m$, original sample $s$, $x_N\sim \mathcal{N}(0,I)$
\begin{algorithmic}[1]
\FOR{$t=N, \ldots, 1$}
\STATE $\epsilon_1, \epsilon_2 \sim \mathcal{N}(0, I)$ if $t>1$, else $\epsilon_1=\epsilon_2=0$
\STATE $y=\sqrt{\bar{\alpha}_t} s+\sqrt{1-\bar{\alpha}_t} \epsilon_1$ if $t>1$, else $s$
\STATE $x_{t-1}=\mu_\theta(x_t, t) + \sigma_\theta(x_t, t)\epsilon_2$ 
\STATE $x_{t-1}=x_{t-1} \odot(1-m)+y \odot m$
\ENDFOR
\RETURN{$x_0$}
\end{algorithmic}
\label{alg:inpaint}
\end{algorithm}

\subsection{External Control (Conditional Generation)}
\label{sec:conditional}
External control means using external signals to condition the generation process.
 We aim to incorporate a general strategy that does not place strong assumptions on the \textit{format} of input control signals. To this end, we use the cross-attention mechanism~\cite{vaswani2017attention} for conditional generation introduced by Latent Diffusion~\cite{rombach2022high} since it is insensitive to the dimension of the condition signals. We also adopted the strategy used by Rombach et al.~\cite{rombach2022high}, which augments the backbone UNet structure with cross-attention layers that map condition signals into the UNet intermediate latent representations.

Formally, to preprocess the external musical signal $c$, we introduce a corresponding encoder $\tau$ that projects $c$ to a latent representation $\tau(c)$. The encoder $\tau$ is pre-trained and fixed during diffusion model training. The cross-attention layers then map $\tau(c)$ to the intermediate layers of the UNet (as shown in Figure~\ref{fig:model}). The conditional training objective is
\begin{equation}
    L_{\text{cond}}(\theta) := \mathbb{E}_{x_0, c, \epsilon, t}\left[\left\|\epsilon-\epsilon_\theta\left(\sqrt{\bar{\alpha}_t} x_0+\sqrt{1-\bar{\alpha}_t}\epsilon, t, \tau(c)\right)\right\|^2\right]
\end{equation}

We use classifier-free guidance (CFG)~\cite{ho2022classifier} to enable both conditioned and unconditioned generation by controlling the intensity of the condition signals during sampling. We refer readers to~\cite{ho2022classifier} and~\cite{dieleman2022guidance} for details on CFG.

\section{Controllable Music Generation} \label{sec:tasks}

\begin{figure*}[!ht]
    \centering
    
    \begin{subfigure}[b]{\textwidth}
    \includegraphics[width=\textwidth]{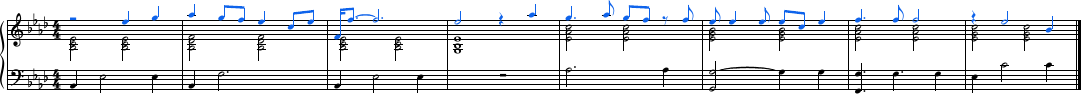}
    \caption{An example of melody generation given accompaniment.}
    \label{fig:staff_mld_gen}
    \end{subfigure}
    
    \begin{subfigure}[b]{\textwidth}
    \includegraphics[width=\textwidth]{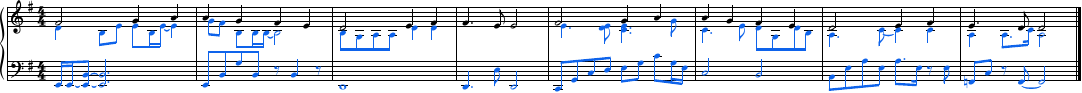}
    \caption{An example of accompaniment generation given melody.}
    \label{fig:staff_acc_gen}
    \end{subfigure}

    \begin{subfigure}[b]{\textwidth}
    \includegraphics[width=\textwidth]{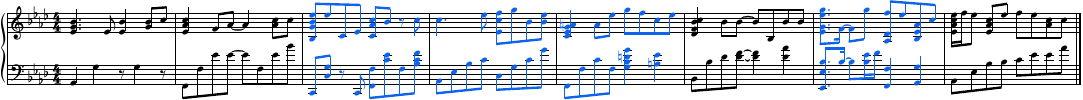}
    \caption{An example of arbitrary segment inpainting.}
    \label{fig:staff_inp_bars}
    \end{subfigure}

    \begin{subfigure}[b]{\textwidth}
    \includegraphics[width=\textwidth]{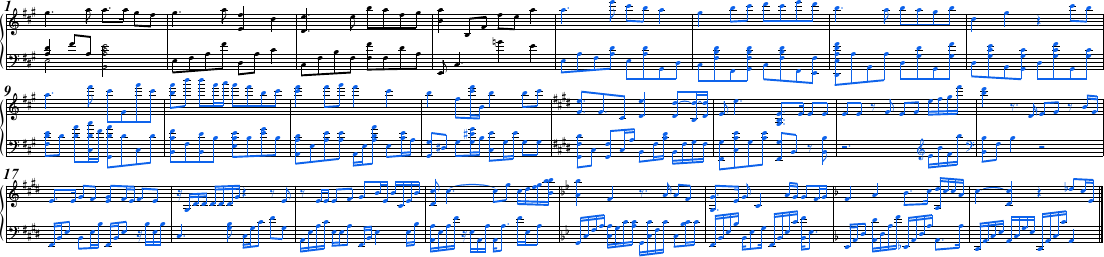}
    \caption{An example of iterative inpainting for long-term music generation.}
    \label{fig:staff_autoreg}
    \end{subfigure}

    \begin{subfigure}[b]{\textwidth}
    \includegraphics[width=\textwidth]{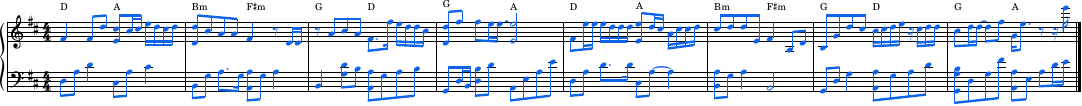}
    \caption{An example of chord-conditioned generation. Chords (annotated above) are used as external condition signals.}
    \label{fig:staff_chd_cond}
    \end{subfigure}

    \begin{subfigure}[b]{\textwidth}
        \includegraphics[width=\textwidth]{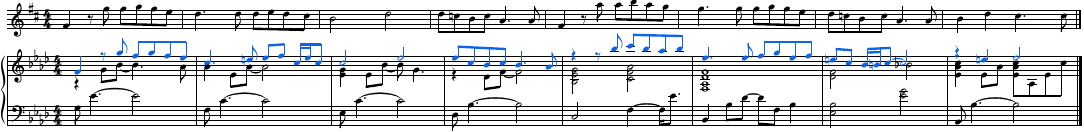}
    \caption{An example of texture-conditioned melody generation. The texture of a given melody (the staff above) is used as external condition signals.}
    \label{fig:staff_txt_mld_gen}
    \end{subfigure}
    \caption{Generated samples in various tasks of controllable music generation. The generated parts are marked in blue. These examples have corresponding hearable demos on the demo page.}
\end{figure*}

In this section, we present four general musical applications our model empowers with internal and external controls: 1) melody generation given accompaniment, 2) accompaniment generation given melody, 3) arbitrary music segment inpainting, and 4) music arrangement given chords or textures. For each application, we provide non-cherry-picked generated samples as a case study. We also refer readers to our \href{https://polyffusion.github.io/}{demo page} for more examples.

\subsection{Melody Generation Given Accompaniment}
\label{sec:inp_mld}
This task is achieved by internal control --- to pre-define the accompaniment part and let the model infill the upper melody. Figure~\ref{fig:staff_mld_gen} shows an example of pop song melody generation given the accompaniment. We see that the melody is consistent with the underlying chords of the given accompaniment, and maintains an overall consistent rhythmic pattern, except for a 16th-note jump at the beginning of the 3rd bar.

\subsection{Accompaniment Generation Given Melody}
Similarly, given a lead melody, we can inpaint its corresponding lower accompaniment. Figure~\ref{fig:staff_acc_gen} shows an example, in which we see that the generated chord sequence suits the key (E minor) of the melody well, realized by a consistent arpeggio texture. The generated counter-melody also fills in the gaps between melody onsets well.

\subsection{Arbitrary Music Segment Inpainting}
The common scenario of music inpainting, also called music infilling~\cite{chang2021variable}, is to generate a music segment that fills in the gap between given past and future contexts. For our model, this task can be fulfilled by masking out the full pitch range of selected bars for inpainting.

Figure~\ref{fig:staff_inp_bars} shows an example of the inpainting process of the 3rd, 4th, 5th, and 7th bars, given the rest as fixed contexts. In the example, the model is capable of generating a full cadence connecting the 7th and the 8th bar, and also a nice applied chord in the non-diatonic progression Gm-Adim-B\musFlat{}m connecting the 5th and the 6th bar.

We also extend the problem setup and let the diffusion model generate \textit{long-term} music by iteratively inpainting the future given the past. Figure~\ref{fig:staff_autoreg} shows an example of a 24-bar generation based on a 4-bar prompt. The model generates 4 bars during each inference and finishes the process with five iterations. We see that the generated music contains a smooth chord progress, with a key modulation towards the end. The long-term textural structure is coherent, however lacking a consistent music theme. 

\subsection{Music Arrangement Given Chords or Textures}
\label{sec:external_control}

Inspired by the \textit{chord}-\textit{texture} disentanglement work~\cite{dai2018music,wang2020learning}, we choose these two factors as the external condition signals for polyphonic generation. In our context, chords refer to the harmonic information, and textures refer to the rhythmic information. The latent chords and textures are encoded using pre-trained VAEs and cross-attended with the backbone UNet.

Beat-wise chords are first extracted by rule-based methods~\cite{pardo2002algorithms,raffel2014mir_eval}, in which we adopted a 36-D chord representation consisting of a 12-D one-hot root encoding, a 12-D one-hot bass encoding and 12-D multi-hot chroma encoding. We then use a chord VAE ~\cite{wang2020learning} to extract a 512-D representation for each 8-bar chord sequence. For texture conditioning, we encode each 2-bar segment with the pre-trained texture encoder in~\cite{wang2020learning} and then concatenate four encoded 256-D representations into a 1024-D vector as an 8-bar texture representation. 

Figure~\ref{fig:staff_chd_cond} demonstrates an example of polyphonic music generation conditioned on chords. In the example, the accompaniment and the melody are mostly chord notes, with a certain degree of non-chord passing and neighboring tones that increase the interestingness of the song.

To show the complex combinations of conditions that the model can handle, we showcase a “texture-specified melody generation” for a given accompaniment segment as an example of the combination of internal and external controls. As shown in Figure~\ref{fig:staff_txt_mld_gen}, We generate the melody part of a given accompaniment segment conditioned on the encoded texture representations of a given melody line. The result preserves a similar rhythmic pattern and fits the tonality of the new accompaniment.

\section{Experiments} \label{sec:exp}

\subsection{Dataset and Training}
\label{sec:train}

We train our model using the POP909 dataset~\cite{wang2020pop909}, a pop song dataset containing around 1K MIDI files. We only keep the pieces with 2/4 and 4/4 meters and cut them into 8-bar music segments with 1-bar hopping size, which results in 64K samples in total. The dataset is randomly split into the training set (90\%) and validation set (10\%) on a song level. The training samples are randomly transposed to all 12 keys for data augmentation.

The classifier-free guidance technique stated in Section~\ref{sec:conditional} combines unconditional and conditional training. We adopt the implementation of DDPM and cross-attention layers in~\cite{labml}. With 1K total diffusion steps, the model converges around 50 epochs (200K steps) on Adam Optimizer~\cite{kingma2014adam} with a constant learning rate 5e-5.

To turn the generated 2-channel piano roll representations into MIDI files, we round them to \{0, 1\} and neglect notes without an onset. In practice, the generation process of 160 8-bar samples report zero invalid notes.

\subsection{Evaluation}

\begin{table*}[h]
\centering
\begin{tabular}{lccccc}
\toprule
                         & (1) Uncond. Gen. & (2) Acc. Gen. & (3) Seg. Inp. & (4) Chord Cond. & (5) Texture Cond. \\ \midrule
Objective Metrics                  & $\mathcal{D}_\mathrm{P}$, $\mathcal{D}_\mathrm{D}$       & $\mathcal{D}_\mathrm{P}$, $\mathcal{D}_\mathrm{D}$                   & $\mathcal{D}_\mathrm{P}$, $\mathcal{D}_\mathrm{D}$  & $\mathcal{D}_\mathrm{P}$, $\mathcal{D}_\mathrm{D}$, CD & $\mathcal{D}_\mathrm{P}$, $\mathcal{D}_\mathrm{D}$, OD         \\
Subjective Metrics & C, N, M & C, N, M, F & N/A & N/A & N/A \\
Generative Length                   & 8 bars       & 8 bars                   & 4 bars & 8 bars & 8 bars            \\
Transformer Baselines & Wang  & Wang              & Chang                                    & Wang & Wang       \\
Sampling Baselines     & Wang          & N/A                     & Wang                                      & Wang  & Wang              \\
\bottomrule
\end{tabular}
\caption{Specifications of the evaluation tasks and the baseline models. C, N, M, F in subjective metrics mean creativity, naturalness, musicality, and fitness respectively. \textit{Wang} refers to the Transformer models (for Transformer baselines) and VAE-based models (for sampling baselines) in~\cite{wang2020learning}; \textit{Chang} refers to the XLNet-based model in~\cite{chang2021variable}.}
\label{tab:eva}

\vspace{5 pt}

\centering
\begin{tabular}{l|ll|ll|ll|lll|lll}
\toprule
               & \multicolumn{2}{l|}{Uncond. Gen.}                          & \multicolumn{2}{l|}{Acc. Gen.}                       & \multicolumn{2}{l|}{Seg. Inp.}                       & \multicolumn{3}{l|}{Chord Cond.}                                      & \multicolumn{3}{l}{Texture Cond.}                                     \\
               & $\mathcal{D}_\mathrm{P}\uparrow$ & $\mathcal{D}_\mathrm{D}\uparrow$ & $\mathcal{D}_\mathrm{P}\uparrow$ & $\mathcal{D}_\mathrm{D}\uparrow$ & $\mathcal{D}_\mathrm{P}\uparrow$ & $\mathcal{D}_\mathrm{D}\uparrow$ & $\mathcal{D}_\mathrm{P}\uparrow$ & $\mathcal{D}_\mathrm{D}\uparrow$ & CD $\downarrow$ & $\mathcal{D}_\mathrm{P}\uparrow$ & $\mathcal{D}_\mathrm{D}\uparrow$ & OD $\downarrow$ \\
               \midrule
Polyffusion    & \textbf{0.89}           & \textbf{0.93}           & \textbf{0.89}           & \textbf{0.96}           & \textbf{0.90}           & \textbf{0.93}           & 0.90                    & \textbf{0.96}           & 0.75            & 0.88                    & \textbf{0.98}           & 1.85            \\
Polyffusion-S5 & \color[HTML]{9B9B9B} 0.89           & \color[HTML]{9B9B9B}0.93           & \color[HTML]{9B9B9B}0.89           & \color[HTML]{9B9B9B}0.96           & \color[HTML]{9B9B9B}0.90           & \color[HTML]{9B9B9B}0.93   & \textbf{0.92}           & 0.81                    & \textbf{0.51}   & 0.87                    & 0.97                    & 1.75            \\
Polyffusion-A  & \color[HTML]{9B9B9B}0.89           & \color[HTML]{9B9B9B}0.93           & \color[HTML]{9B9B9B}0.89           & \color[HTML]{9B9B9B}0.96           & \color[HTML]{9B9B9B}0.90           & \color[HTML]{9B9B9B}0.93                             & 0.90                    & 0.94                    & 0.79            &    \textbf{0.95}                   &   \textbf{0.98}                      &  4.37               \\
Transformer & 0.78                    & 0.84                    & 0.88                    & 0.89                    & \textbf{0.90}           & 0.83                    & 0.87                    & 0.88                    & 0.56            & 0.84                    & 0.93                    & \textbf{0.13}   \\
Sampling       & 0.86                    & 0.90                    & N/A & N/A                            & 0.89                    & 0.91                    & 0.86                    & 0.90                    & 0.70            & 0.91           & 0.93                    & 0.20            \\ \bottomrule
\end{tabular}
\caption{The objective evaluation and ablation study results. The statistics of generation, accompaniment generation and segment inpainting are identical for three Polyffusion models (hence gray-out for the latter two models) since they share the same internal control method.}
\label{tab:result}
\end{table*}

To validate the generation quality and control effectiveness of our model, we conducted both objective and subjective evaluations on 5 tasks: (1) unconditional generation, (2) accompaniment generation, (3) segment inpainting, (4) chord-conditioned generation, and (5) texture-conditioned generation. Tasks 2-3 focus on the evaluation of internal controls, and tasks 4-5 focus on external controls. Table~\ref{tab:eva} summarizes the evaluation method for each task. 

\subsubsection{Evaluation Metrics}
\textbf{Objective metrics}: To objectively measure the music quality for all 5 tasks, we use the averaging overlapped area of pitch distribution ($\mathcal{D}_\mathrm{P}$) and duration distribution ($\mathcal{D}_\mathrm{D}$) from~\cite{ren2020popmag}, which measure the distribution similarity of pitch and duration between the generated samples and ground truth. Additionally, we introduce \textit{chord distance} (CD)~\cite{ren2020popmag} and \textit{onset distance} (OD) to evaluate the efficacy of external control. These metrics measure the $\ell_2$ distance of chord (for task 4) and onset distribution (for task 5) between the generated samples and the chord/texture condition.%

\textbf{Subjective metrics}: Subjective metrics include \textit{creativity} (C), \textit{naturalness} (N), and \textit{musicality} (M), which provide a perceptive evaluation complementing the objective musical quality metrics. To demonstrate the efficacy of internal control, we pick accompaniment generation as an example and add a \textit{fitness} (F) metric to evaluate how well the generated parts fit in with the given melody.

\subsubsection{Baseline models}

We use two types of models as our baselines:

\textbf{Transformer models}: As suggested in the polyphonic representation disentanglement study~\cite{wang2020learning}, applying a Transformer on disentangled latent codes yields better results than raw token predictions. Following~\cite{wang2020learning}, we train a Transformer to predict the chord and texture representations from melody representations. For unconditional generation (task 1), we sample the latent spaces of the first 2-bar melody and then predict its accompaniment and the following content. For accompaniment generation (task 2) and external conditioning (tasks 4-5), the melody (task 2), chord (task 4), or texture (task 5) latent representation is directly encoded as the condition for the Transformer. We adopt the XLNet-based model proposed in~\cite{chang2021variable} for the music segment inpainting task (task 3).
    
\textbf{Sampling-based models}: We adopt the VAE-based disentanglement model in~\cite{wang2020learning} and generate music segments by sampling the latent spaces. For unconditional generation (task 1), we sample from the chord and texture latent spaces of the first and the last 2 bars, then linearly interpolate the middle latent codes to form a coherent 8-bar segment. For inpainting (task 3), we also use linear interpolation on latent codes to infill the missing bars. For external conditioning (tasks 4-5), the chord (task 4) or texture (task 5) latent component is directly encoded from the given condition.

\subsection{Comparative Results}

We calculate the average of each objective metric on 160 generated samples for each task. As shown in Table~\ref{tab:result}, Polyffusion and its variations achieve the highest objective scores in tasks 1-4. For controllability, our model yields competitive results on segment inpainting and chord-conditioned generation. For the texture-conditioned generation task, our model does not perform as well as the baseline but is capable of preserving the general musical texture, since the baseline model is explicitly trained on texture reconstruction targets, while the texture condition of our model only serves as a hint for the generation.

We also show the effectiveness of classifier-free guidance in Table~\ref{tab:result}. With a guidance scale of $5$, the model (Polyffusion-S5) shows improved controllability on both chord conditioning and texture conditioning. Notably, a large guidance scale for chord conditions negatively impacts the $\mathcal{D}_\mathrm{D}$ metric. We speculate that this is because notes regular in length provide clearer chord context, which can be noticed in the guidance demos.
\begin{figure}[t]
  \centering
    \includegraphics[width=0.46\textwidth]{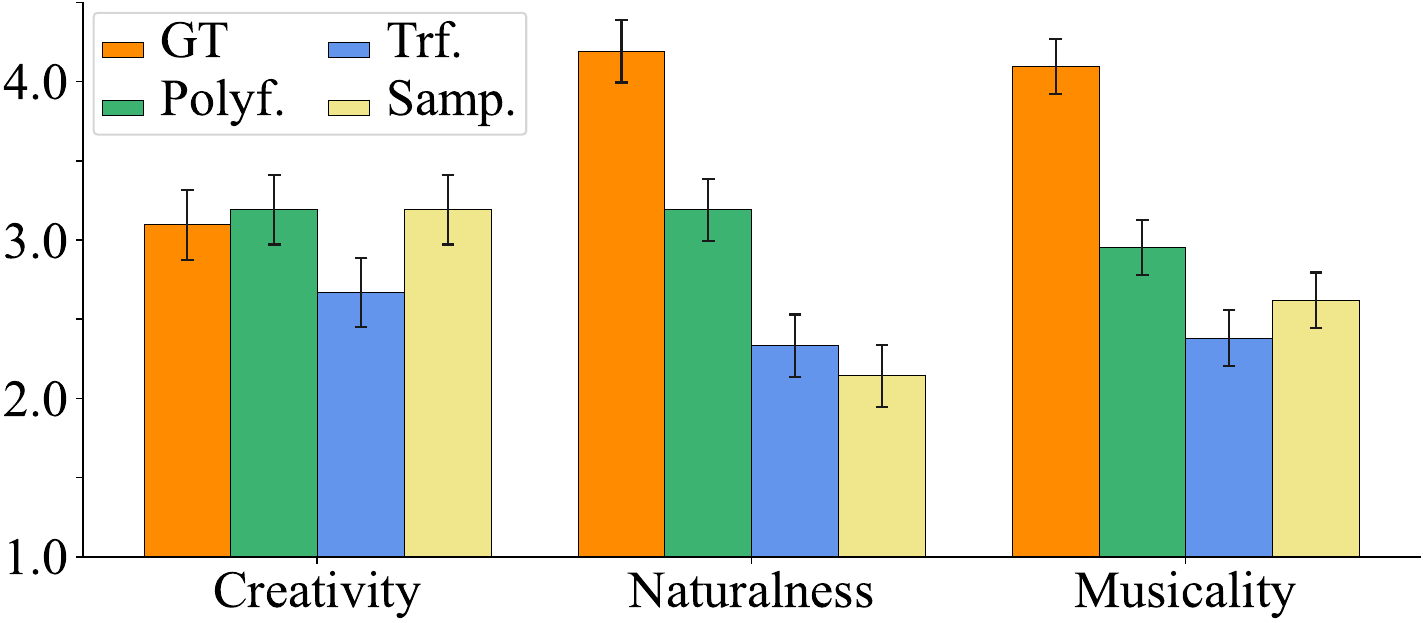}
  \caption{Subjective evaluation for unconditional generation.}
  \label{subjective_unconditional}
\end{figure}

\begin{figure}[t]
  \centering
    \includegraphics[width=0.48\textwidth]{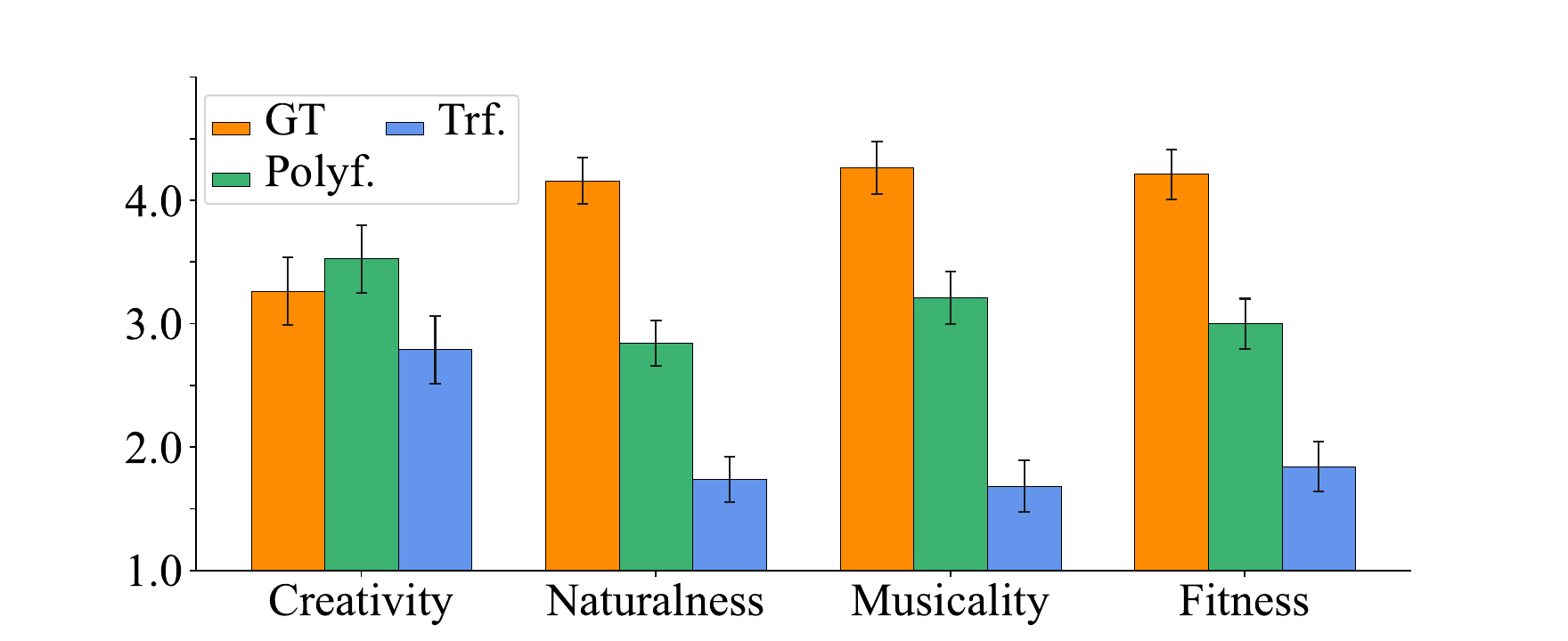}
  \caption{Subjective evaluation for accompaniment generation.}
  \label{subjective_accompaniment}
\end{figure}

For subjective evaluation, we invite participants to rate the generation quality via a double-blind online survey. Our survey consists of 4 groups of samples of unconditional generation and accompaniment generation, respectively. Each group contains a ground-truth piece, generated samples by Polyffusion and all baselines with random orders. 36 participants completed our survey. Each participant rated 4 random groups on average based on a 5-point scale. The evaluation results are shown in Figure~\ref{subjective_unconditional} and \ref{subjective_accompaniment}. The height of each bar represents the mean rating, and the error bars are MSEs computed by within-subject ANOVA \cite{scheffe1999analysis}. We report a significantly better performance (p-value $<$ 0.05) of Polyffusion than baseline models in \textit{naturalness} and \textit{musicality} for both tasks and in \textit{fitness} for accompaniment generation. Interestingly, Polyffusion even outperforms the ground truth on the \textit{creativity} metric.

\subsection{Ablation Study}

We perform an ablation test on the use of VAE encoders for condition signals. For both chord conditioning and texture conditioning, we remove the corresponding pre-trained encoders. The ablated model of chord conditioning uses concatenated 36-D chord vectors as the condition signals. The ablated model of texture conditioning uses a modified piano roll representation~\cite{wang2020learning}. Both models are trained with the same settings as the proposed model. Table~\ref{tab:result} shows that the ablated models (Polyffusion-A) perform worse than the proposed models on the controllability metrics (CD \& OD), showing the advantage of using disentangled latent representations as condition signals for diffusion models.

\section{Conclusion and Future Work}

In this paper, we propose a diffusion model for polyphonic symbolic music generation. We show that an image-like piano roll representation is effective for modeling the musical context for a high-quality score generation. We specify two methods for controllable generation: internal control via masked generation, and external control via conditioning using cross-attention. Experiments show that our method achieves higher quality and controllability compared to the Transformer and sampling-based baselines on both internal and external control tasks.%

We regard the diffusion framework as a prospective direction for future work on controllable music generation, since it achieves fine-grained controls over high-quality generation and enables a wide spectrum of arrangement applications. Currently, our generation is limited to quantized music scores without performance features. We plan to extend this methodology to expressive performance modeling. Several new controls can also be introduced to facilitate human-AI co-creation of symbolic music, e.g., hierarchical structure controls (e.g., music segment labels) and multimodal controls (e.g., text descriptions).

\bibliography{ISMIRtemplate}

\end{document}